\documentclass[12pt]{article}
\usepackage{amsmath}
\usepackage{amssymb}
\usepackage{amsthm}

\begin{document}       
\title{Asymmetric Simple Exclusion Process and 
Modified Random Matrix Ensembles}
\author{
Taro Nagao
{\footnote {\tt e-mail: nagao@math.nagoya-u.ac.jp}}
~~and Tomohiro Sasamoto
{\footnote {\tt e-mail: sasamoto@stat.phys.titech.ac.jp}}
}
\date{}
\maketitle

\begin{center}
\it 
$^*$Graduate School of Mathematics, Nagoya University, \\ Chikusa-ku, 
Nagoya 464-8602, Japan
\end{center}
\begin{center}
\it
$^{\dag}$Department of Physics, Tokyo Institute of Technology, \\
Oh-okayama 2-12-1, Meguro-ku, Tokyo 152-8551, Japan
\end{center}

\begin{abstract}
We study the fluctuation properties of the asymmetric simple 
exclusion process (ASEP) on an infinite one-dimensional 
lattice. When $N$ particles are initially situated in 
the negative region with a uniform density $\rho_-=1$, 
Johansson showed the equivalence of the current fluctuation 
of ASEP and the largest eigenvalue distribution of random 
matrices. We extend Johansson's formula and derive 
modified ensembles of random matrices, corresponding 
to general ASEP initial conditions. Taking the scaling 
limit, we find that a phase change of the asymptotic current 
fluctuation occurs at a critical position. 
\end{abstract}

\maketitle
[Keywords: asymmetric exclusion process, random matrices]

\newpage

\newtheorem{theorem}{Theorem}
\newtheorem{proposition}[theorem]{Proposition}
\newtheorem{corollary}[theorem]{Corollary}
\newtheorem{lemma}[theorem]{Lemma}
\def\a{\alpha}
\def\b{\beta}
\def\Ai{{\rm Ai}}
\def\d{\delta}
\def\e{\epsilon}
\def\g{\gamma}
\def\G{\Gamma}
\def\i{\infty}
\def\l{\lambda}
\def\bra{\langle}
\def\ket{\rangle}
\def\t{\tau}
\def\Z{\mathbb{Z}}
\def\p{\varphi}
\def\s{\sigma}
\def\P{\mathbb{P}}
\def\i{\infty}
\def\dd#1#2{\frac{d #1}{d #2}}
\def\ddd#1#2{\frac{d^2 #1}{d #2^2}}

The one-dimensional asymmetric simple exclusion process 
(ASEP) is one of the most transparently analyzable models 
in non-equilibrium statistical mechanics 
\cite{Liggett1985,Liggett1999,Spohn1991,Schuetz2000}. In 
the last two decades, many exact results have been 
worked out for the stationary and dynamical behavior 
of ASEP. Among all, it is now known to belong to the 
Kardar-Parisi-Zhang (KPZ) universality 
class \cite{KPZ1986}. 
\par
The study of the fluctuation properties of ASEP 
has been recently renewed after Johansson's 
work\cite{Johansson2000}. Johansson evaluated 
the current fluctuation of ASEP for an initial 
condition with a step at the origin, using the 
connections to random matrix theory\cite{Mehta1991} 
by way of combinatorics of semistandard 
Young tableaux\cite{Stanley1999}. The key 
ingredient of Johansson's argument is an integral 
formula, which evidently shows the equivalence of 
the current fluctuation and the largest eigenvalue 
distribution of random matrices. 
\par
Following Johansson's work and relying on 
Baik and Rains' result\cite{BR2000,BR2001a,BR2001b} 
on combinatorics, Pr\"ahofer and Spohn\cite{PS2002} 
made a plausible conjecture for a phase diagram of 
the fluctuation properties. In their scheme, the 
universality classes of random matrices characterize 
the ASEP fluctuation phases. 
\par
In this paper, we derive a generalization of 
Johansson's integral formula, which enables 
us to treat general initial conditions 
and establishes a connection to modified 
ensembles of random matrices. An asymptotic 
analysis of this new formula yields a phase 
diagram in agreement with Pr\"ahofer and Spohn's 
conjecture. 
\par   
Our approach is based on the determinantal 
formula for the Green function found 
by Sch\"utz\cite{Schuetz1997b}. Suppose that $N$ 
particles labelled $1,2,\cdots,N$ from the left 
start from $y_1,y_2,\ldots,y_N$ at time $0$.
The probability that particles are at $x_1,x_2,\ldots,x_N$ at time $t$
is given by
\begin{align}
\label{green}
&\quad P(x_1,x_2,\cdots,x_N;t|y_1,y_2,\cdots,y_N;0) \notag\\
&= 
\det[F_{k-j}(x_k-y_j;t)]_{j,k=1}^N,
\end{align}
where
\begin{equation}
  F_n(m;t)
  =
  {\rm e}^{-t} \frac{t^m}{m!} \sum_{k=0}^{\infty} 
  \frac{(n)_k}{(m+1)_k}\frac{t^k}{k!}.
\end{equation}
Here $(n)_k=n(n+1)\cdots (n+k-1)$. 
The function $F_n(m,t)$ has some interesting properties 
such as 
\begin{align}
 \label{int_F}
 \int_0^{t} {\rm d}s F_n(m;s) 
 &= 
 F_{n+1}(m+1;t)-F_{n+1}(m+1;0) , \\
 \label{sum_F}
 \sum_{l=m}^{\i} F_n(l;t) 
 &= 
 F_{n+1}(m;t).
\end{align}
It is also remarked that $F_n(m,t)$ can be written in terms of the 
confluent hypergeometric function (or Laguerre function),
though it is not explicitly stated in \cite{Schuetz1997b}.

The Green function is already a nontrivial result. However, 
in order to further study physical quantities, one still has to 
take summations over initial and final configurations.
Our first observation is that a certain summation 
also results in a determinantal 
expression: 
\begin{proposition}
Suppose that $N$ particles labelled $1,2,\cdots,N$ from the left 
start from $y_1,y_2,\ldots,y_N$ at time $0$.
The probability that the particle $j$ ($1\leq j \leq N$) 
has hopped at least $X_j-y_j$ steps before time $t$ is given by
\begin{align}
\label{sum_det}
&\quad \sum_{X_j \leq x_j \, (1\leq j \leq N)} 
P(x_1,x_2,\cdots,x_N;t|y_1,y_2,\cdots,y_N;0) \notag\\
&= 
\det[F_{k-j+1}(X_k-y_j;t)]_{j,k=1}^N.
\end{align}
\end{proposition}
\noindent
This can be proved by using (\ref{sum_F}) successively. 
Notice that the only difference between 
(\ref{green}) and (\ref{sum_det}) is the suffix of $F_n$:  
in the latter it is bigger than that in the former by one.

In this article we are interested in the fluctuation of 
the integrated current or the position of a particle.
For this purpose, one only needs the special case of (\ref{sum_det}) 
with $X_j-y_1 = M+j-1$ ($1\leq j \leq N$), which we
state as a corollary.
\begin{corollary}
Suppose that $N$ particles start from $y_1,y_2,\ldots,y_N$ at time $0$.
The probability that the leftmost particle has hopped at least 
$M$ steps before time $t$ is given by
\begin{align}
\label{sum_det_sp}
&\quad\P[x_1 \geq y_1+M] \notag\\
&=
\sum_{y_1+M\leq x_1<x_2<\cdots < x_N} 
P(x_1,x_2,\cdots,x_N;t|y_1,y_2,\cdots,y_N;0) \notag\\
&= 
\det[F_{k-j+1}(y_1-y_j+M+k-1;t)]_{j,k=1}^N.
\end{align}
\end{corollary}

Next we derive another formula for the same quantity.
To obtain the result, we need 
\begin{lemma}
\label{int}
If $f(t_N,t_{N-1},\cdots,t_1)$ is a totally  antisymmetric 
function of $N$ variables, the following holds.
\begin{align}
&\quad 
\int_0^t {\rm d}t_N \prod_{j=1}^{N-1}\int_0^{t_{N-j+1}} {\rm d}t_{N-j}
\prod_{k=1}^{N-1}\int_0^{t_{N-k}} {\rm d}s_{N-k,1}
\notag\\ &\times
\prod_{l=2}^k \int_0^{s_{N-k,l-1}} {\rm d}s_{N-k,l}
f(t_N,s_{N-1,1},s_{N-2,2},\cdots,s_{1,N-1}) \notag\\
&=
\prod_{j=0}^{N-1}\int_0^t {\rm d}t_{N-j}
\prod_{k=1}^{N-1}\int_0^{t_{N-k}} {\rm d}s_{N-k,1}
\prod_{l=2}^k \int_0^{s_{N-k,l-1}} {\rm d}s_{N-k,l}
\notag\\ &\times
f(t_N,s_{N-1,1},s_{N-2,2},\cdots,s_{1,N-1}) \notag\\
&=
\frac{1}{\prod_{j=1}^N j!}
\prod_{j=0}^{N-1}\int_0^t {\rm d}t_{N-j}
\prod_{1\leq j<k\leq N} (t_k-t_j)  
f(t_N,\cdots,t_1).
\label{prop2}
\end{align}
\end{lemma}

\vspace{3mm}\noindent
{\bf Proof.}
Using a mathematical induction on $N$, we first
prove a slightly generalized version of the first equality 
\begin{align}
\label{a_sym}
&\quad   
\int_0^t {\rm d}t_N \int_0^{t_N} {\rm d}t_{N-1} \cdots \int_0^{t_2} 
{\rm d}t_1
\notag\\ &\times
\int_0^{t_{N-1}} {\rm d}s_{N-1,1} 
\int_0^{t_{N-2}} {\rm d}s_{N-2,1} \int_0^{s_{N-2,1}} {\rm d}s_{N-2,2}
\cdots \notag\\
&\times   
\int_0^{t_1} {\rm d}s_{1,1} \int_0^{s_{1,1}} {\rm d}s_{1,2} \cdots 
\int_0^{s_{1,N-2}} {\rm d}s_{1,N-1}
\notag\\ &\times
f(t_N,s_{N-1,1},s_{N-2,2},\cdots,s_{1,N-1};t_1) \notag\\
&=
\int_0^t {\rm d}t_N \int_0^{t} {\rm d}t_{N-1} \cdots \int_0^{t} {\rm d}t_1
\notag\\ &\times
\int_0^{t_{N-1}} {\rm d}s_{N-1,1} 
\int_0^{t_{N-2}} {\rm d}s_{N-2,1} \int_0^{s_{N-2,1}} {\rm d}s_{N-2,2}
\cdots \notag\\
&\times   
\int_0^{t_1} {\rm d}s_{1,1} \int_0^{s_{1,1}} {\rm d}s_{1,2} \cdots 
\int_0^{s_{1,N-2}} {\rm d}s_{1,N-1}
\notag\\ &\times
f(t_N,s_{N-1,1},s_{N-2,2},\cdots,s_{1,N-1};t_1),
\end{align}
in which the function $f$ also depends on $t_1$.

When $N=1$, the statement is trivial.

Now we assume that the statement is true for $N-1$
and show that the statement is true for $N$.
If we divide the region of the integration over $t_1$ on the right hand side (RHS) 
as $\int_0^t {\rm d}t_1 = \int_0^{t_2} {\rm d}t_1 +\int_{t_2}^t {\rm d}t_1$, 
the term corresponding to $\int_{t_2}^t {\rm d}t_1$ is rewritten as
\begin{align}
&\quad   
\int_0^t {\rm d}t_N \int_0^t {\rm d}t_{N-1} \cdots \int_0^t {\rm d}t_2 
\int_{t_2}^t {\rm d}t_1
\notag\\ &\times
\int_0^{t_{N-1}} {\rm d}s_{N-1,1} 
\int_0^{t_{N-2}} {\rm d}s_{N-2,1} \int_0^{s_{N-2,1}} {\rm d}s_{N-2,2}
\cdots 
\notag\\ &\times
\int_0^{t_2} {\rm d}s_{2,1} \int_0^{s_{2,1}} {\rm d}s_{2,2} \cdots 
\int_0^{s_{2,N-3}} {\rm d}s_{2,N-2} 
\notag\\ &\times
\int_0^{t_1} {\rm d}s_{1,1} \int_0^{s_{1,1}} {\rm d}s_{1,2} \cdots 
\int_0^{s_{1,N-2}} {\rm d}s_{1,N-1} 
\notag\\ &\times
f(t_N,s_{N-1,1},s_{N-2,2},\cdots,s_{2,N-2},s_{1,N-1};t_1) \notag\\
&=
\int_0^t {\rm d}t_N \int_0^t {\rm d}t_{N-1} \cdots \int_0^t {\rm d}t_3 \int_0^t {\rm d}t_1 
\notag\\ &\times 
\int_0^{t_{N-1}} {\rm d}s_{N-1,1} 
\int_0^{t_{N-2}} {\rm d}s_{N-2,1} \int_0^{s_{N-2,1}} {\rm d}s_{N-2,2}
\cdots 
\notag\\ &\times   
\int_0^{t_1} {\rm d}t_2\int_0^{t_2} {\rm d}s_{2,1} \int_0^{s_{2,1}} {\rm d}s_{2,2} \cdots 
\int_0^{s_{2,N-3}} {\rm d}s_{2,N-2} 
\notag\\ &\times
\int_0^{t_1} {\rm d}s_{1,1} \int_0^{s_{1,1}} {\rm d}s_{1,2} \cdots 
\int_0^{s_{1,N-2}} {\rm d}s_{1,N-1} 
\notag\\ &\times
f(t_N,s_{N-1,1},s_{N-2,2},\cdots,s_{2,N-2},s_{1,N-1};t_1).
\end{align}
This is zero since $f$ is antisymmetric in $s_{2,N-2}$ and 
$s_{1,N-1}$ while they are integrated in the same way.  
Hence the region $[0,t]$ of the integration over $t_1$ on the RHS 
of (\ref{a_sym}) can be replaced by $[0,t_2]$. One obtains  
\begin{align}
\label{step1}
&\quad \text{the RHS of (\ref{a_sym})} \notag\\
&=
\int_0^t {\rm d}t_N \int_0^t {\rm d}t_{N-1} \cdots \int_0^t {\rm d}t_2 
\notag\\ &\times
\int_0^{t_{N-1}} {\rm d}s_{N-1,1} 
\int_0^{t_{N-2}} {\rm d}s_{N-2,1} \int_0^{s_{N-2,1}} {\rm d}s_{N-2,2}
\cdots 
\notag\\ &\times   
\int_0^{t_2} {\rm d}s_{2,1} \int_0^{s_{2,1}} {\rm d}s_{2,2} \cdots 
\int_0^{s_{2,N-3}} {\rm d}s_{2,N-2} 
\notag\\ &\times   
~g(t_N,s_{N-1,1},s_{N-2,2},\cdots,s_{2,N-2};t_2), 
\end{align}
where
\begin{align}
&\quad 
g(t_N,s_{N-1,1},\cdots,s_{2,N-2};t_2) \notag\\
&\equiv
\int_0^{t_2} {\rm d}t_1 \int_0^{t_1} {\rm d}s_{1,1} \int_0^{s_{1,1}} {\rm d}s_{1,2} \cdots 
\int_0^{s_{1,N-2}} {\rm d}s_{1,N-1} 
\notag\\ &\times
f(t_N,s_{N-1,1},s_{N-2,2},\cdots,s_{2,N-2},s_{1,N-1};t_1).
\end{align}
We note that $g(t_N,s_{N-1,1},\cdots,s_{2,N-2};t_2)$ is 
antisymmetric in $N-1$ variables $t_N,s_{N-1,1},\cdots,s_{2,N-2}$. 
From the assumption of mathematical induction,
$\int_0^t {\rm d}t_N \int_0^t {\rm d}t_{N-1} \cdots \int_0^t {\rm d}t_2 $
in (\ref{step1}) can be replaced by
$\int_0^t {\rm d}t_N \int_0^{t_N} {\rm d}t_{N-1} \cdots \int_0^{t_3} {\rm d}t_2 $.
Using the definition of $g$, we arrive at the desired statement.

Next the second equality of the lemma 
is proved by a mathematical induction on $N$.

When $N=1$, the statement is trivial.

Let us assume that the statement is true for $N-1$
and show that the statement is true for $N$.
Let us first define
\begin{align}
&\quad h(t_N,s_{N-1,1},\cdots,s_{2,N-2}) \notag\\
&\equiv
\int_0^t {\rm d}t_1 \int_0^{t_1} {\rm d}s_{1,1} \int_0^{s_{1,1}} {\rm d}s_{1,2} \cdots 
\int_0^{s_{1,N-2}} {\rm d}s_{1,N-1}
\notag\\ &\times
f(t_N,s_{N-1,1},s_{N-2,2},\cdots,s_{1,N-1}).
\end{align}
This function is antisymmetric in $N-1$ variables 
$t_N,s_{N-1,1},s_{N-2,2}, \cdots,s_{2,N-2}$. 
It can be further rewritten as
\begin{align}
&\quad h(t_N,s_{N-1,1},\cdots,s_{2,N-2}) \notag\\
&=
\int_0^t {\rm d}s_{1,N-1} \int_{s_{1,N-1}}^t {\rm d}t_1 
\int_{s_{1,N-1}}^{t_1} {\rm d}s_{1,1} 
\int_{s_{1,N-1}}^{s_{1,1}} {\rm d}s_{1,2} \cdots 
\int_{s_{1,N-1}}^{s_{1,N-3}} {\rm d}s_{1,N-2} 
\notag\\ &\times
f(t_N,s_{N-1,1},s_{N-2,2},\cdots,s_{2,N-2},s_{1,N-1}) \notag\\
&=
\frac{1}{(N-1)!} \int_0^t {\rm d}s_{1,N-1} (t-s_{1,N-1})^{N-1} 
f(t_N,s_{N-1,1},s_{N-2,2},\cdots,s_{2,N-2},s_{1,N-1}) \notag\\
&=
\frac{1}{(N-1)!} \int_0^t {\rm d}t_1 (t-t_1)^{N-1} 
f(t_N,s_{N-1,1},s_{N-2,2},\cdots,s_{2,N-2},t_1).
\label{ht} 
\end{align}
Using the assumption of mathematical induction, 
we rewrite the middle expression of (\ref{prop2}) as 
\begin{align}
&\quad
\int_0^t {\rm d}t_N \int_0^{t} {\rm d}t_{N-1} \cdots \int_0^{t} {\rm d}t_1
\notag\\ &\times
\int_0^{t_{N-1}} {\rm d}s_{N-1,1} 
\int_0^{t_{N-2}} {\rm d}s_{N-2,1} \int_0^{s_{N-2,1}} {\rm d}s_{N-2,2}
\cdots 
\notag\\ &\times   
\int_0^{t_2} {\rm d}s_{2,1} \int_0^{s_{2,1}} {\rm d}s_{2,2} \cdots 
\int_0^{s_{2,N-3}} {\rm d}s_{2,N-2}
\notag\\ &\times   
\int_0^{t_1} {\rm d}s_{1,1} \int_0^{s_{1,1}} {\rm d}s_{1,2} \cdots 
\int_0^{s_{1,N-2}} {\rm d}s_{1,N-1} 
\notag\\ &\times
f(t_N,s_{N-1,1},s_{N-2,2},\cdots,s_{2,N-2},s_{1,N-1}) \notag\\
&=
\int_0^t {\rm d}t_N \int_0^{t} {\rm d}t_{N-1} \cdots \int_0^{t} {\rm d}t_2
\notag\\ &\times   
\int_0^{t_{N-1}} {\rm d}s_{N-1,1} 
\int_0^{t_{N-2}} {\rm d}s_{N-2,1} \int_0^{s_{N-2,1}} {\rm d}s_{N-2,2}
\cdots 
\notag\\ &\times  
\int_0^{t_2} {\rm d}s_{2,1} \int_0^{s_{2,1}} {\rm d}s_{2,2} \cdots 
\int_0^{s_{2,N-3}} {\rm d}s_{2,N-2}
\notag\\ &\times  
~h(t_N,s_{N-1,1},s_{N-2,2},\cdots,s_{2,N-2}) 
\notag\\ &=
\frac{1}{\prod_{j=1}^{N-1} j!}
\int_0^t {\rm d}t_N \int_0^{t} {\rm d}t_{N-1} \cdots \int_0^{t} {\rm d}t_2
\prod_{2\leq j<k\leq N} (t_k-t_j)  
\notag\\ &\times  
~h(t_N,t_{N-1},\cdots,t_2). 
\end{align} 
From (\ref{ht}), we find that it is equal to
\begin{align}
&\quad   
\frac{N}{\prod_{j=1}^{N} j!}
\int_0^t {\rm d}t_N \int_0^{t} {\rm d}t_{N-1} \cdots \int_0^t {\rm d}t_1
\prod_{2\leq j<k\leq N} (t_k-t_j)  ~(t-t_1)^{N-1} 
\notag\\ &\times
f(t_N,t_{N-1},\cdots,t_1)
\notag\\ &=
\frac{1}{\prod_{j=1}^{N} j!}
\int_0^t {\rm d}t_N \int_0^{t} {\rm d}t_{N-1} \cdots \int_0^t {\rm d}t_1
\sum_{m=1}^N (-1)^{m-1} 
\notag\\ &\times
\prod_{\substack{1\leq j<k\leq N\\j,k\neq m}} (t_k-t_j)  ~(t-t_m)^{N-1} 
f(t_N,t_{N-1},\cdots,t_1).
\end{align}
Thus the second equality of the proposition is established because of 
the identity 
\begin{equation}
\sum_{m=1}^N (-1)^{m-1} 
\prod_{\substack{1\leq j<k\leq N\\j,k\neq m}} (t_k-t_j)  ~(t-t_m)^{N-1} 
=
\prod_{1\leq j<k\leq N} (t_k-t_j).
\end{equation}
This identity can be proved by observing that the left hand 
side is antisymmetric polynomials in $t_1,t_2,\cdots,t_N$ and 
that the coefficient of $t_1^{N-1}$ on both sides is 
$(-1)^{N-1} \prod_{2\leq j<k\leq N}(t_k-t_j)$.
\qed

Now we state one of the main results of this paper:
\begin{theorem}
\begin{align}
\label{main}
&\quad\P[x_1 \geq y_1+M] 
\notag\\
&= 
\frac{1}{\prod_{j=1}^N j!} 
\prod_{j=0}^{N-1}\int_0^t {\rm d}t_{N-j}
\prod_{1 \leq j < k \leq N} (t_k - t_j) 
f(t_N,\cdots,t_1),
\end{align}
where
\begin{align}
\label{LUEgen}
&\quad  
f(t_N,t_{N-1},\cdots,t_1) \notag\\
&= 
\det[F_{-j+1}(y_1-y_j+M-1;t_{N-k+1})]_{j,k=1}^N.
\end{align}
When $M\leq y_N-y_1$, the RHS of (\ref{main}) 
should be understood as its analytic 
continuation with respect to $M$.

\end{theorem}

\vspace{3mm}\noindent
{\bf Proof.} 
First let us assume that the condition $M > y_N-y_1$ is satisfied.
Applying Lemma \ref{int} to the RHS of 
(\ref{sum_det_sp}), we have
\begin{align}
&\quad \sum_{y_1+M\leq x_1<x_2<\cdots < x_N} 
P(x_1,x_2,\cdots,x_N;t|y_1,y_2,\cdots,y_N;0) 
\notag\\
&=   \int_0^t {\rm d}t_N \int_0^{t_N} {\rm d}t_{N-1} \cdots \int_0^{t_2} 
{\rm d}t_1
\notag\\ &\times 
\det[ F_{k-j}(y_1-y_j+M+k-2;t_{N-k+1})]_{j,k=1}^N
\notag\\
&= \int_0^t {\rm d}t_N \int_0^{t_N} {\rm d}t_{N-1} \cdots \int_0^{t_2} {\rm d}t_1
\notag\\ &\times
   \int_0^{t_{N-1}} {\rm d}s_{N-1,1} 
   \int_0^{t_{N-2}} {\rm d}s_{N-2,1} \int_0^{s_{N-2,1}} {\rm d}s_{N-2,2}
   \cdots 
\notag\\ &\times 
\int_0^{t_1} {\rm d}s_{1,1} \int_0^{s_{1,1}} {\rm d}s_{1,2} \cdots 
\int_0^{s_{1,N-2}} {\rm d}s_{1,N-1}
\notag\\ &\times
f(t_N,s_{N-1,1},s_{N-2,2},\cdots,s_{1,N-1})
\notag\\
&= 
\frac{1}{\prod_{j=1}^N j!} 
\prod_{j=0}^{N-1}\int_0^t {\rm d}t_{N-j}
\prod_{1 \leq j < k \leq N} (t_k - t_j) 
f(t_N,\cdots,t_1).
\end{align}

Next we would like to consider the case $M\leq y_N-y_1$.
Assume, however, $M>y_N-y_1$ for a moment.
Let us denote by $G_N(t)$ the RHS of (\ref{main}) with $f$ 
replaced by $f_0$, which is defined by
\begin{align}
\label{f0}
&\quad  
f_0(t_N,t_{N-1},\cdots,t_1) \notag\\
&= 
\det[F_0(y_1-y_j+M-1;t_{N-k+1})]_{j,k=1}^N.
\end{align}
By expanding the determinants, one obtains 
\begin{align}
\label{Ggamma}
 G_N(t)
 &=
 \sum_{\s,\t} \text{sgn} (\s\t) 
 \frac{\g(y_1-y_{\t(N)}+M+\s(1)-1,t)}{\G(y_1-y_{\t(N)}+M)} 
 \times \cdots \notag\\
 &\quad \times
 \frac{\g(y_1-y_{\t(1)}+M+\s(N)-1,t)}{\G(y_1-y_{\t(1)}+M)}.
\end{align}
Here $\s$ and $\t$ are permutations of $1,2,\cdots,N$.
The function $\g(x,t)$ is the incomplete Gamma function 
defined by
\begin{equation}
 \g(x,t) = \int_0^t {\rm e}^{-s} s^{x-1} {\rm d}s,
\end{equation}
where the real part of $x$ is positive. Just as in the case 
of the ordinary Gamma function $\Gamma(x) = \g(x,\infty)$, 
the incomplete Gamma function can be analytically continued 
via the contour integral representation
\begin{equation}
 \g(x,t) = \frac{1}{{\rm e}^{2\pi i x}-1} \int_{C(t)} 
{\rm e}^{-z} z^{x-1} {\rm d}z,
\end{equation}
where $C(t)$ is the contour starting from $t$, enclosing the 
origin anticlockwise and ending at $t$. This expression is 
meaningful for all $x$ on the complex plane except at the 
locations of poles. In fact it is known that $\g(x,t)$ has 
an expression,
\begin{equation}
 \g(x,t) = \sum_{n=0}^{\i} \frac{(-1)^n t^{x+n}}{n!(x+n)},
\end{equation}
and hence has precisely the same singularities as $\G(x)$; 
the poles are located at $x=0,-1,-2,\cdots$ and the corresponding
 residues are $1,-1,1/2,\cdots$. Taking these facts into consideration, 
one realizes that the RHS of (\ref{Ggamma}) can be analytically 
continued to the region $M\leq y_N-y_1$, because each (incomplete or 
ordinary) gamma function can be analytically continued and the 
singularities of the gamma functions on the numerators are cancelled 
by those on the denominators.

By using (\ref{sum_F}), one can rewrite $f$ as a linear 
combination of $f_0$ and vice versa. Hence the RHS of 
(\ref{main}) can also be analytically continued
to the region $M \leq y_N-y_1$. 
The value for an integer $M$ is finite and gives the 
desired probability $\P[x_1 \geq y_1+M]$.
\qed

\medskip
For the special case $y_j=-N+j$ ($j=1,2,\cdots,N$), 
the RHS of (\ref{main}) reduces to 
\begin{align}
\label{LUE}
\frac{1}{Z_{M,N}} \prod_{j=0}^{N-1}\int_0^t {\rm d}t_{N-j} 
\prod_{1\leq j<k\leq N} (t_k-t_j)^2 \prod_{j=1}^N (t_j^{M-N} {\rm e}^{-t_j}),
\end{align}
where $Z_{M,N} = \prod_{j=1}^N j! (M-N+j-1)!$. 
This is the Johansson's integral formula\cite{Johansson2000}, 
the RHS of which yields the largest eigenvalue 
distribution of the Laguerre unitary ensemble of 
random matrices. One can therefore analyze the 
ASEP current fluctuation by means of random matrix 
theory. Johansson derived this formula by way of 
combinatorics of the semistandard Young tableaux. 
The derivation of our theorem is physically more 
transparent, because we have just summed up 
the Green function. Furthermore our theorem generalizes 
Johansson's formula and allows us to study arbitrary 
initial conditions. In addition, the RHS of (\ref{main}) 
can be considered as a modification of matrix integrals 
appearing in the random matrix theory \cite{Mehta1991}.

The initial condition in the above theorem has 
a fixed configuration of $N$ particles.
One can also consider a superposition of 
initial conditions by summing up over particle 
configurations. In fact, it can be even 
technically easier to work with a superposition.
Notice that $G_N(t)$ defined in the proof of the 
above theorem represents 
the probability that the leftmost particle has hopped at least 
$M$ steps before time $t$ for a superposition of initial 
conditions. The technical advantage of replacing 
$f$ by $f_0$ is that it can be written in a form
\begin{equation}
\label{detcd}
G_N(t) =  \prod_{l=0}^{N-1}\int_0^t {\rm d}t_{N-l} 
 \det\left[c_j(t_k)\right]
 \det\left[d_j(t_k)\right]
 \prod_{l=1}^N w(t_l).
\end{equation}
When $M\leq y_N-y_1$, the RHS should be again 
understood as special cases of its analytic 
continuation with respect to $M$. This point will 
be discussed in some detail afterwards.

To proceed further,
we construct monic polynomials $C_j(t)$ and $D_j(t)$, so that 
$C_j(t)$ ($D_j(t)$) is a linear combination of 
$c_k(t)$ ($d_k(t)$) with $k=0,1,\cdots, j$ and satisfy
\begin{equation}
\label{CDortho}
\bra C_j,D_k \ket \equiv  \int_0^{\i} C_j(t) D_k(t) w(t) 
{\rm d}t  = h_j \d_{jk}, \ \ \ j,k=0,1,2,\cdots,N-1,
\end{equation}
where $h_j$ ($j=0,1,\cdots,N-1$) is a constant.
After some reflection, one sees that 
$C_j$ and $D_j$ can be expressed as 
\begin{equation}
\label{Cj}
 C_j(t)
 \propto
 \begin{vmatrix}
  c_0(t) & \bra c_0,d_0 \ket & \cdots & \bra c_0,d_{j-1} \ket \\
  c_1(t) & \bra c_1,d_0 \ket & \cdots & \bra c_1,d_{j-1} \ket \\
  \vdots & \vdots            &        & \vdots \\ 
  c_j(t) & \bra c_j,d_0 \ket & \cdots & \bra c_j,d_{j-1} \ket 
 \end{vmatrix},
\end{equation}
\begin{equation}
\label{Dj}
 D_j(t)
 \propto
 \begin{vmatrix}
  d_0(t) & \bra d_0,c_0 \ket & \cdots & \bra d_0,c_{j-1} \ket \\
  d_1(t) & \bra d_1,c_0 \ket & \cdots & \bra d_1,c_{j-1} \ket \\
  \vdots & \vdots            &        & \vdots \\ 
  d_j(t) & \bra d_j,c_0 \ket & \cdots & \bra d_j,c_{j-1} \ket 
 \end{vmatrix},
\end{equation}
with the condition that they are monic (the coefficients 
of $c_j(t)$ and $d_j(t)$ are $1$).

Let us now define
\begin{align}
 \phi_j(t) &= \frac{1}{\sqrt{h_j}}C_j(t) \sqrt{w(t)}, \\
 \psi_j(t) &= \frac{1}{\sqrt{h_j}}D_j(t) \sqrt{w(t)}, 
\end{align}
so that (\ref{CDortho}) reads
\begin{equation}
 \label{phipsiortho}
 \int_0^{\i} \phi_j(t) \psi_k(t) {\rm d}t = \d_{jk}.
\end{equation}
Then we can show that 
$G_N(t)$ has an expression
\begin{equation}
 \label{GNdet}
  G_N(t) = \det(1-K_N),
\end{equation}
where the determinant on the RHS is the Fredholm determinant
and the kernel $K_N(x,y) \chi_I(y)$ is given by
\begin{equation}
 K_N(x,y) = \sum_{j=0}^{N-1} \phi_j(x) \psi_j(y),
\end{equation}
\begin{equation}
\chi_I(x) = \left\{ \begin{array}{ll} 1, & x \in I, \\ 
0, & {\rm otherwise} \end{array} \right.
\end{equation}
and $I = (t,\infty)$. This Fredholm determinant expression 
can be derived by following the strategy of 
Gaudin\cite{Gaudin1961}.

In the following, we consider a special choice
\begin{equation} 
\label{ychoice}
 y_1=-N+1-Y, \ \ y_j=-N+j~~(j=2,3,\cdots,N)
\end{equation}
in (\ref{f0}), where $Y$ is a nonnegative integer.
This choice is one of the simplest generalizations of the 
case studied by Johansson. The corresponding initial condition 
is that the particles are distributed in the negative region 
with a uniform density
\begin{equation}
\rho_-=\frac{N}{Y+N}
\end{equation}
and that the particle density in the positive region is 
zero. As we noted, the fluctuation was already conjectured by 
Pr\"ahofer and Spohn\cite{PS2002}. Our formula allows us to 
prove their conjecture. In terms 
of $c_j(t)$ and $d_j(t)$, the choice (\ref{ychoice}) 
corresponds to 
\begin{align}
\label{cj_sp}
 c_j(t) &= t^j, \quad (j=0,1,\cdots,N-1), \\
\label{dj_sp}
 d_j(t) &= \begin{cases}
	    t^j,  & (j=0,1,\cdots, N-2), \\
            t^K,  & j=N-1,
	   \end{cases} \\
\label{w_sp}
 w(t)   &= t^{\a} {\rm e}^{-t},
\end{align}
with $K=N+Y-1$. For the case specified 
by (\ref{cj_sp})-(\ref{w_sp}), 
$C_j(t)$ is the monic version of the Laguerre 
polynomial,
\begin{equation}
\label{CjLj}
 C_j(t) = k_j^{-1} L_j^{(\a)}(t), \quad (j=0,1,\cdots, N-1),
\end{equation}
with $k_j = (-1)^j/j!$. In terms of $C_j(t)$'s, the 
orthogonality of the Laguerre 
polynomials reads
\begin{equation}
\label{Cortho}
 \int_0^{\i} C_j(t) C_k(t) w(t) {\rm d}t = h_j^{(0)} \d_{j,k},
\end{equation}
with
$h_j^{(0)} = j! \G(M-K+j)$.
The polynomial $D_j(t)$ can also be equated with the monic 
Laguerre polynomial for $j=0,\cdots,N-2$,
\begin{equation}
 D_j(t) = k_j^{-1} L_j^{(\a)}(t), \quad (j=0,1,\cdots, N-2).
\end{equation}
Hence, for $j=0,1,\cdots,N-2$, $h_j$ is identical to $h_j^{(0)}$:
\begin{equation}
 h_j = h_j^{(0)} = j! \G(M-K+j), \quad (j=0,\cdots, N-2).
\end{equation}
As for $D_{N-1}(t)$, one uses the determinant formula (\ref{Dj}).
After some computation, one obtains 
\begin{equation}
\label{DNm1}
 D_{N-1}(t) = \G(M) \sum_{j=N-1}^K \frac{(-K)_j}{\G(M-K+j)} L_j^{(\a)}(t).
\end{equation}
The explicit formulas for $C_{N-1}(t)$ and $D_{N-1}(t)$ allow 
one to compute $h_{N-1}$ in (\ref{CDortho}). The result is 
\begin{equation}
 h_{N-1} = (-1)^{N-1}\G(M)(-K)_{N-1}.
\end{equation}

As we mentioned below (\ref{detcd}), for the 
case $M \leq y_N-y_1$, some care should be 
taken for the interpretation of the formulas appearing 
in the above discussion: the orthogonality relation 
(\ref{CDortho}) has its meaning as its analytic 
continuation. For $j=0,1,\cdots,N-1$ and $k=0,1,\cdots,N-2$, 
(\ref{CDortho}) is equivalent to the orthogonality relation 
of the Laguerre polynomials,
\begin{equation}
 \frac{\G(j+1)}{\G(j+\a+1)}
 \int_0^{\i} L_j^{(\a)}(t) L_k^{(\a)}(t) t^{\a} {\rm e}^{-t} {\rm d}t
 =
 \d_{j,k}.
\label{Lortho}
\end{equation}
Using a formula
\begin{equation}
 L_j^{(\a)}(t) 
 = 
 \sum_{l=0}^j \frac{(-1)^l \G(j+\a+1)}{\G(j-l+1)\G(\a+l+1)}
              \frac{t^l}{\G(l+1)},
\end{equation}
one finds
\begin{align}
 &\quad
 \frac{\G(j+1)}{\G(j+\a+1)}
 \int_0^{\i} L_j^{(\a)}(t) L_k^{(\a)}(t) t^{\a} {\rm e}^{-t} {\rm d}t 
 \notag\\
 &=
 \sum_{m=0}^j\sum_{n=0}^k
 \frac{(-1)^{m+n}\G(j+1)\G(k+\a+1)\G(m+n+\a+1)}
      {m!n!~\G(j-m+1)\G(\a+m+1)\G(k-n+1)\G(\a+n+1)}.
\end{align}
As a function of $\a$, there is no singularity on the RHS.

For $j=0,1,\cdots,N-1$ and $k=N-1$, on the other hand, one 
has to use the expression (\ref{DNm1}) of $D_{N-1}$. 
However, since only the $j=N-1$ term in (\ref{DNm1}) 
gives a nonzero contribution, the orthogonality relation 
(\ref{CDortho}) can be analytically continued in a 
similar manner. 

Finally we consider the scaling limit of the current fluctuation 
for the special case (\ref{ychoice}) with a uniform initial 
density $\rho_-$ in the negative region. The results are given 
in terms of the transcendental functions ${\cal F}_2(s)$ 
and ${\cal F}_1(s)$, which were originally introduced in 
random matrix theory. The largest 
eigenvalue distribution of Gaussian (or Laguerre) 
unitary and orthogonal ensembles are written 
in terms of ${\cal F}_2(s)$ and ${\cal F}_1(s)$, respectively. 

If one introduces the function $u(x)$ which satisfies
the Painlev\'e II equation,
\begin{equation}
 \frac{\partial^2}{\partial x^2} u = 2 u^3 + xu, 
\end{equation}
with the asymptotics
\begin{equation}
 u(x) \sim \Ai(x) \quad x\to\i,
\end{equation}
the functions ${\cal F}_2(s)$ and ${\cal F}_1(s)$ are 
represented as\cite{TW1994,TW1996}
\begin{align}
 {\cal F}_2(s) &= \exp \left[-\int_s^{\i} (x-s) u(x)^2 {\rm d}x\right], \\
 {\cal F}_1(s) &= \exp \left[-\frac12\int_s^{\i} (x-s) u(x)^2 
{\rm d}x-\frac12\int_s^{\i}u(x) {\rm d}x \right].
\end{align}
They also have the Fredholm determinant representations\cite{Forrester1993,
Forrester2000p}
\begin{align}
 {\cal F}_2(s) &= \det(1-K_2), \\
 ({\cal F}_1(s))^2 &= \det(1-K_1).
\end{align}
The kernels $K_2(\xi,\eta) \chi_I(\eta)$ and $K_1(\xi,\eta) \chi_I(\eta)$ 
, respectively, are given by
\begin{align}
 K_2(\xi,\eta)
 &=
 K_{\rm Airy}(\xi,\eta) 
 = 
 \frac{\Ai(\xi) \Ai'(\eta) - \Ai(\eta) \Ai'(\xi)}{\xi-\eta}, \\
 \label{K1}
 K_1 (\xi,\eta)
 &=
 K_{\rm Airy}(\xi,\eta) + \Ai(\xi) \int_0^{\i} \Ai(\eta-\l) {\rm d}\l
\end{align}
and $I = (s,\infty)$.
\par
Now let us consider the asymptotics of the current fluctuation.
We set $M=Y+\g N$ ($\g > 0$). This corresponds to measuring 
the current at a position $\a=(\g-1)N$. The results depend 
on the initial density on the negative sites, $\rho_-$.  Now 
the results of our asymptotic analysis (in agreement with 
Pr\"ahofer and Spohn's conjecture\cite{PS2002}) are 
summarized as
\begin{theorem}
 Set $t=cN+dN^{1/3}s$ where
 \begin{equation}
  \label{cd_def}
  c = (1+\sqrt{\g})^2, \quad d=  \g^{-1/6}(1+\sqrt{\g})^{4/3}.
 \end{equation}
 (i)For $\lim_{N \rightarrow \infty} Y/N < \sqrt{\g}$,
 $G_N(t)$ tends to ${\cal F}_2(s)$ in the limit $N\to\i$;
 \begin{equation}
  \lim_{N\to\i} G_N(t) = {\cal F}_2(s).
 \end{equation}
 (ii)When $\lim_{N \rightarrow \infty} Y/N = \sqrt{\g}$ (i.e. $\rho_-=1/(1+\sqrt{\g})$, 
$\a = N (1 - 2 \rho_-)/\rho_-^2$), 
$G_N(t)$ tends to ${\cal F}_1(s)^2$ in the limit $N\to\i$;
 \begin{equation}
  \lim_{N\to\i} G_N(t) = ({\cal F}_1(s))^2.
 \end{equation}
\end{theorem}

\vspace{3mm}\noindent
{\bf Proof.} 
Let us rewrite $K_N(x,y)$ as
\begin{align}
 K_N(x,y) 
 &=
 \sum_{j=0}^{N-1}\phi_j^{(0)}(x)\phi_j^{(0)}(y) + 
 \phi_{N-1}(x) \psi_{N-1}(y) 
 \notag\\
 &\quad -  \phi_{N-1}^{(0)}(x) \phi_{N-1}^{(0)}(y) ,
\label{KNdevide}
\end{align}
where $\phi_j^{(0)}(x)=\phi_j(x) (j=0,\cdots,N-2),
\phi_{N-1}^{(0)}(x)=C_{N-1}(x)\sqrt{w(x)}/\sqrt{h_{N-1}^{(0)}}$.
Using the Christoffel-Darboux formula for $C_j$,
the first term is rewritten as
\begin{align}
 \sqrt{(M-Y)N} \frac{ \phi_N^{(0)}(x) \phi_{N-1}^{(0)}(y)
                     -\phi_N^{(0)}(y) \phi_{N-1}^{(0)}(x)}{x-y}.
\end{align}
Here and in the following we need the asymptotic behavior of the 
Laguerre polynomials. In terms of the function defined by 
\begin{equation}
 \p_n^{(\a)}(x) 
 =
 (-1)^n \left(\frac{n!}{(\a+n)!}\right)^{1/2} \sqrt{w(x)}L_n^{(\a)}(x),
\end{equation}
one has
\begin{equation}
\label{p_n_a}
 \p_{n+\frac{d\l\sqrt{\g}}{c} n^{1/3}}^{((\g-1)n)}(x) 
 \sim
 \frac{1+\sqrt{\g}}{\g^{1/4} d n^{1/3}}\Ai(\xi-\l) 
\end{equation}
($x = c N + d N^{1/3} \xi$), which can be obtained by 
applying the saddle point method
to the integral representation of $\p_n^{(\a)}(x)$.
Then it is not difficult to see that the first term in 
(\ref{KNdevide}) multiplied by $dN^{1/3}$ tends to 
$K_{\rm Airy}(\xi,\eta)$.

We next consider the second term on the RHS of (\ref{KNdevide}).  
One sees that, when $\lim_{N \rightarrow \infty} Y/N < 
\sqrt{\g}$, the term with $j=N-1$ becomes dominant in the 
limit $N\to\i$.  Using this expression and the asymptotics of 
$\p$ in (\ref{p_n_a}), we can show that the second and 
the third terms in (\ref{KNdevide}) are negligible.
Hence the case $(i)$ is proved.

When $Y=\sqrt{\g} N$, one uses the asymptotics, (\ref{p_n_a}),
of $\p$ to find
\begin{align}
 dN^{1/3}\phi_{N-1}(x) \psi_{N-1}(y) 
 \sim
 \Ai(\xi) \int_0^{\i} {\rm d}\l \Ai(\eta-\l).
\end{align}
The third term in (\ref{KNdevide}) is again negligible.
This leads us to $K_1(\xi,\eta)$ in (\ref{K1}), ending the 
proof of the case $(ii)$.
\qed

\par
\medskip

The density profile of ASEP in the scaling limit 
$t \rightarrow \infty$ (with $N = t/c$) is 
known to be\cite{Liggett1999}
\begin{equation}
\rho(x) = \left\{ \begin{array}{cc} \rho_-, & 
x < (1 - 2 \rho_-)t, \\
(1 - (x/t))/2, &  
(1 - 2 \rho_-)t < x < t, \\
0, & x > t. \end{array} \right.
\end{equation}
We can therefore conclude that the current 
fluctuation is described by ${\cal F}_2(s)$ in the 
region with a linearly decreasing density. 
At the singular position 
$x_c = (1 - 2 \rho_-) t$, a phase change occurs (a 
Gaussian fluctuation is conjectured\cite{PS2002} 
at $x < (1 - 2 \rho_-) t$). 

Let us finally remark that, owing to the particle-hole 
symmetry of ASEP, we can extend the result to a 
measurement at $-\a$. In that case, the initial 
particle density in the negative region is fixed to $1$. 

To summarize, we have studied the fluctuation properties of 
the asymmetric simple exclusion process (ASEP) on an 
infinite one-dimensional lattice. We first obtained a 
determinantal formula by summing up the Green function. 
This can be turned into a form of a modified random matrix 
ensemble. As an application of our formula, we considered the 
fluctuation of the integrated current when the particles 
are initially distributed in the negative region with a 
uniform density $\rho_-$. A generalization of the technique 
in random matrix theory has been utilized to study the 
asymptotic properties of the model. Consequently we 
found a phase change of the current fluctuation at a 
critical position.

The authors would like to thank Prof. Makoto Katori, Prof. 
Hideki Tanemura and Mr. Takashi Imamura for useful discussions 
and comments.


\begin{thebibliography}{10}

\bibitem{Liggett1985}
T.~M. Liggett,
\newblock {\em Interacting Particle Systems}
\newblock (Springer-Verlag, 1985).

\bibitem{Liggett1999}
T.~M. Liggett,
\newblock {\em Stochastic Interacting Systems: Contact, Voter, and Exclusion
  Processes}
\newblock (Springer-Verlag, 1999).

\bibitem{Spohn1991}
H.~Spohn,
\newblock {\em Large Scale Dynamics of Interacting Particles}
\newblock (Springer, 1991).

\bibitem{Schuetz2000}
G.~M. Sch{\"u}tz,
\newblock Exactly solvable models for many-body systems far from equilibrium, 
\newblock in: C.~Domb and J.~L. Lebowitz (editors)
{\em Phase Transitions and Critical Phenomena 19} (2000). 

\bibitem{KPZ1986}
M. Kardar, G. Parisi and Y. C. Zhang,
\newblock Phys. Rev. Lett. 56 (1986) 889.

\bibitem{Johansson2000}
K.~Johansson, 
\newblock Commun. Math. Phys. 209 (2000) 437.

\bibitem{Mehta1991}
M.~L. Mehta,
\newblock {\em Random Matrices}
\newblock (Academic, 2nd edition, 1991).

\bibitem{Stanley1999}
R.~P. Stanley,
\newblock {\em Enumerative Combinatorics 2}
\newblock (Springer, 1999).

\bibitem{BR2000}
J.~Baik and E.~M. Rains, 
\newblock J. Stat. Phys. 100 (2000) 523.

\bibitem{BR2001a}
J.~Baik and E.~M. Rains,
\newblock Duke Math. J. 109 (2001) 1.

\bibitem{BR2001b}
J.~Baik and E.~M. Rains,
\newblock Duke Math. J. 109 (2001) 205.

\bibitem{PS2002}
M.~Pr{\"a}hofer and H.~Spohn,
\newblock in: V.~Sidoravicius (editor) {\em In and out of equilibrium, vol. 51
  of \it Progress in Probability} (2002) pp. 185--204.

\bibitem{Schuetz1997b}
G.~M. Sch{\"u}tz,
\newblock J. Stat. Phys. 88 (1997) 427.

\bibitem{Gaudin1961}
M. Gaudin, 
\newblock Nucl. Phys. 25 (1961) 447.

\bibitem{TW1994}
C.~A. Tracy and H.~Widom,
\newblock Commun. Math. Phys. 159 (1994) 151.

\bibitem{TW1996}
C.~A. Tracy and H.~Widom,
\newblock Commun. Math. Phys. 177 (1996) 727.

\bibitem{Forrester1993}
P.~J. Forrester,
\newblock Nucl. Phys. B 402 (1993) 709.

\bibitem{Forrester2000p}
P.~J. Forrester,
\newblock Painlev{\'e} transcendent evaluation of the scaled distribution of
  the smallest eigenvalue in the Laguerre orthogonal and symplectic ensembles, 
nlin.SI/0005064.


\end{thebibliography}
\end{document}